\documentclass[prd,twoside,preprintnumbers,superscriptaddress,twocolumn,nofootinbib]{revtex4}

\usepackage{amsmath,graphicx,slashed,multirow,color}

\begin{document}

\preprint{CERN-PH-TH/2014-147}

\title{Dark matter direct detection constraints from gauge bosons loops}

\author{Andreas Crivellin}
\email{andreas.crivellin@cern.ch}
\affiliation{CERN Theory Division, CH-1211 Geneva 23, Switzerland}
\author{Ulrich Haisch}
\email{u.haisch1@physics.ox.ac.uk}
\affiliation{CERN Theory Division, CH-1211 Geneva 23, Switzerland}
\affiliation{Rudolf Peierls Centre for Theoretical Physics, University of Oxford,
OX1 3PN Oxford, United Kingdom}

\date{\today}

\begin{abstract}
While many interactions of dark matter~(DM) with the standard model~(SM) affect direct detection and LHC searches, there are only a few operators generating annihilation of DM into photons. All of these operators, except four of them, give rise to unsuppressed rates, rendering indirect detection superior to other search strategies. For two of the four effective interactions with velocity-suppressed  annihilation cross sections, we identify a new type of loop effect which significantly enhances the  associated direct detection rates. As a result, relevant constraints from next-generation direct detection  experiments on DM-SM interactions, so far only bounded by the LHC, are obtained. 
 \end{abstract}
 
\maketitle

\section{Introduction}
\label{sec:1}

In quantum field theory the weakness of interactions is often related to the fact that the operators that induce the couplings are irrelevant. The ``darkness'' of DM could therefore be a natural consequence  of the DM states having only higher-dimensional interactions with photons, and more generally with electroweak~(EW) gauge bosons. In our article we will consider $SU(2)_L \times U(1)_Y$ gauge-invariant operators up to dimension~7.

We start with scalar DM. In the case of real scalar DM, one has the following dimension-6  operators 
\begin{equation} \label{eq:1}
\chi^2 \, B_{\mu \nu}  B^{\mu \nu} \,, \qquad
\chi^2 \, W_{\mu \nu}^i  W^{i, \mu \nu} \,,
\end{equation}
as well as terms where the $U(1)_Y$ or $SU(2)_L$ field strength tensor $B_{\mu \nu} = \partial_\mu B_\nu - \partial_\nu B_\mu$ or  $W_{\mu \nu}^i = \partial_\mu W_\nu^i - \partial_\nu W_\mu^i + g_2 \hspace{0.25mm} \epsilon^{ijk}  \hspace{0.25mm}  W_\mu^j \hspace{0.25mm} W_\mu^k$  is replaced by its dual $\tilde B_{\mu \nu} = 1/2 \, \epsilon_{\mu \nu \lambda \rho} \hspace{0.25mm} B^{\lambda \rho}$ or  $\tilde W_{\mu \nu}^i = 1/2 \, \epsilon_{\mu \nu \lambda \rho} \hspace{0.25mm} W^{i, \lambda \rho}$. All of these operators lead to unsuppressed annihilation into $\gamma\gamma$ and $\gamma Z$~(see~e.g.~\cite{Rajaraman:2012db,Frandsen:2012db,Rajaraman:2012fu}) so that the bounds from direct detection or colliders are not competitive with the indirect search limits. The same statements apply if DM is a complex scalar.

Turning to fermionic DM, the leading effects arise in the Dirac case from dimension-5 operators of electric or magnetic dipole type. Explicitly, one has the term 
\begin{equation} \label{eq:2}
\bar \chi \hspace{0.25mm} \sigma_{\mu \nu} \chi \, B^{\mu \nu} \,,
\end{equation}
and its dual counterpart (in the case of real Wilson coefficients). Here  $\sigma_{\mu\nu}=i/2 \, (\gamma_\mu\gamma_\nu-\gamma_\nu\gamma_\mu)$. As pointed out for instance in~\cite{Barger:2010gv,Banks:2010eh}, such operators give rise to long-range tree-level interactions between DM and nucleons so that the resulting direct detection constraints are by far too stringent to allow observable $\gamma$-ray signals. 

For a Dirac fermion, the following dimension-7 operators exist 
\begin{equation} \label{eq:3}
\begin{split}
\bar \chi \hspace{0.25mm} \sigma_{\mu \nu} \chi \, B^\mu_{\ \alpha} \tilde B^{\nu \alpha} \,, \qquad 
\bar \chi \hspace{0.25mm} \sigma_{\mu \nu} \chi \, W^{i, \mu}_{\ \alpha} \hspace{0.25mm} \tilde W^{i, \nu \alpha} \,,
\\[2mm]
\bar \chi \hspace{0.25mm} \sigma_{\mu \nu} \chi \, B^{\mu \nu} \, \phi^\dag \phi \,, \qquad 
\bar \chi \hspace{0.25mm} \sigma_{\mu \nu} \chi \, W^{i, \mu \nu} \, \phi^\dag \tau^i \phi \,.
\end{split}
\end{equation}
Here $\phi$ denotes the SM Higgs doublet and $\tau^i$ are the usual $SU(2)_L$  generators. Contributions involving two Higgs fields $\phi$ and a dual field strength tensor are also possible. The whole set of these operators induces unsuppressed annihilation rates to either $\gamma h$ or $\gamma \gamma$ and $\gamma Z$~\cite{Rajaraman:2012db,Rajaraman:2012fu}, rendering  indirect detection bounds superior to the other DM search strategies.  

Irrespectively of whether DM is Dirac or Majorana, in addition the operators 
\begin{equation} \label{eq:4}
O_B= \bar \chi \chi \, B_{\mu \nu} B^{\mu \nu } \,, \qquad 
O_W= \bar \chi \chi \, W_{\mu \nu}^i W^{i, \mu \nu } \,,
\end{equation}
can be written down~(cf.~\cite{Weiner:2012cb,Weiner:2012gm}). While they induce velocity-suppressed annihilation cross sections, their counterparts with a pseudo-scalar DM current lead to unsuppressed annihilation rates~\cite{Rajaraman:2012db,Frandsen:2012db,Rajaraman:2012fu}. Operators of the form~(\ref{eq:4}) involving $\tilde B_{\mu \nu}$ or $\tilde W_{\mu \nu}^i$ do not produce direct detection signals even at the one-loop level~\cite{Frandsen:2012db} and we thus do not consider such contributions in what follows.  

Since indirect detection provides no meaningful constraint,  the  interactions~(\ref{eq:4}) are, up to dimension~7, the only  DM-photon couplings for which  loop-induced direct detection may be phenomenologically relevant.  While  the  one-loop contributions to direct detection originating from operators with electromagnetic field strength tensors have been studied~\cite{Frandsen:2012db}, such a computation is, to the best of our knowledge, not available for~$O_B$ and~$O_W$. The main goal of this work is to close this gap. Like in previous works~\cite{Frandsen:2012db,Weiner:2012cb,Hisano:2010ct,Hisano:2011cs,Hill:2011be,Hisano:2012wm,Haisch:2013uaa,Hill:2013hoa,Hill:2014yka,Crivellin:2014qxa}, we find that also for the terms~(\ref{eq:4}) virtual exchange of SM particles can significantly change the predictions for the DM-nucleon scattering cross section. In~Sec.~\ref{sec:2} we spell out the necessary ingredients to come to this conclusion. In~Sec.~\ref{sec:3} we review the restrictions arising from the invisible decay $h \to \bar \chi \chi$ of the SM Higgs boson, from missing energy~($\slashed{E}_T$) searches at the LHC and from the DM relic density. We present our numerical results in Sec.~\ref{sec:4}, before concluding in Sec.~\ref{sec:5}.

\section{Loop-induced direct detection}
\label{sec:2}

Below we present  a concise discussion of how SM loops arising from insertions of the effective operators $O_B$ and~$O_W$ give rise to DM scattering with nuclei. In order to calculate the corresponding cross section one has to perform the following four separate steps: $i)$ renormalisation group~(RG) evolution from the new-physics~(NP) scale~$\Lambda$, where the  interactions~(\ref{eq:4}) are generated, down to the EW scale $\mu_w$; $ii)$ computation of matching corrections at the EW scale obtained by integrating out the top quark, the Higgs, the $Z$ and the $W$ boson; $iii)$ RG evolution from the EW scale to the hadronic scale $\mu_l \simeq 1 \, {\rm GeV}$, taking into account heavy quark thresholds; $iv)$ calculation of the nucleon matrix elements of all operators that are present at $\mu_l$, including those induced by operator mixing.   

\begin{figure}
\begin{center}
\includegraphics[height=26ex]{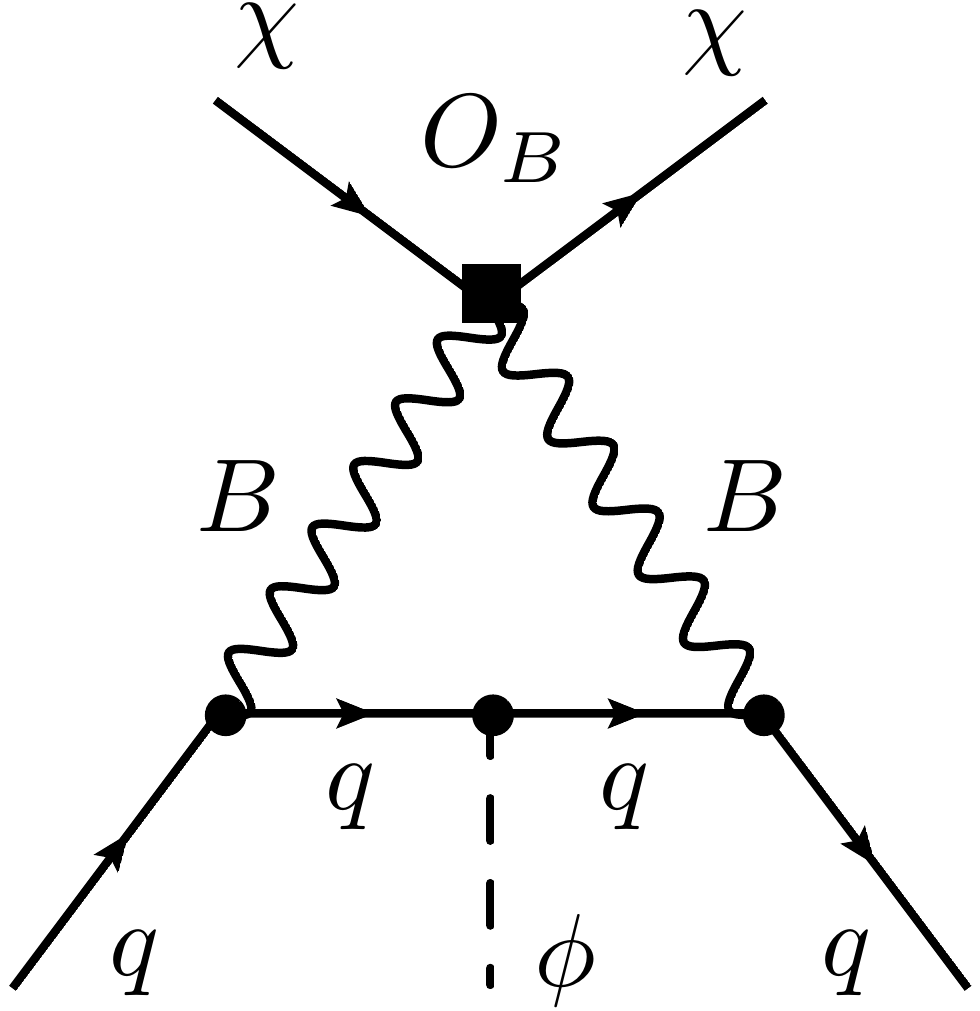}  \qquad  
\includegraphics[height=27ex]{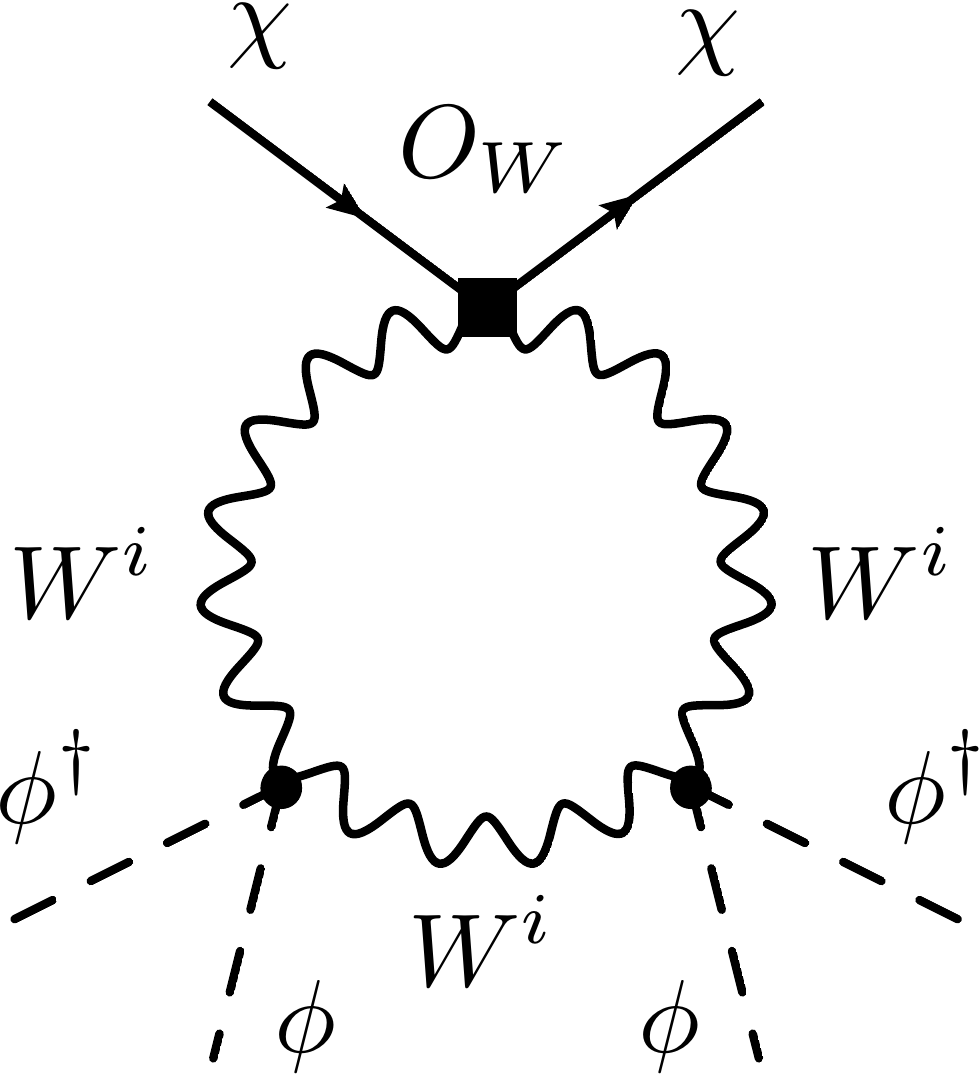}
\end{center}
\vspace{-2mm}
\caption{\label{fig:1} Examples of Feynman diagrams  generating  mixing of  $O_B$ into $O_y$~(left) and $O_W$ into $O_\phi$~(right). The operator insertions are indicated by squares while the  $SU(2)_L \times U(1)_Y$ dimension-4 vertices are represented by dots.}
\end{figure}

It turns out that to perform step $i)$ one has to consider besides~(\ref{eq:4}), the two additional dimension-7 operators 
\begin{equation} \label{eq:5} 
O_y = y_q \bar \chi \chi \, \bar q \hspace{0.25mm} \phi \hspace{0.25mm} q\,, \qquad
O_\phi = \bar \chi \chi  \left ( \phi^\dag \phi \right)^2 \,.
\end{equation}
Here $y_q = \sqrt{2} \hspace{0.25mm} m_q/v$  are the SM quark Yukawa couplings and $v \simeq 246 \, {\rm GeV}$ denotes the vacuum expectation value~(VEV) of the Higgs doublet $\phi$. The effective Lagrangian relevant for scales $\mu$ with $\Lambda > \mu > \mu_w$ is hence 
\begin{equation} \label{eq:6}
{\cal L}_{\rm eff} = \sum_{k=B,W,y,\phi} \frac{C_k (\mu)}{\Lambda^3} \, O_k \,. 
\end{equation}
The Feynman graphs in the unbroken $SU(2)_L \times U(1)_Y$ theory that generate~(\ref{eq:5}) are shown in Fig.~\ref{fig:1}. From these diagrams we obtain in the leading logarithmic~(LL) approximation the following corrections 
\begin{equation} \label{eq:7}
\begin{split}
C_y (\mu_w) & \simeq \frac{3 \hspace{0.25mm} Y_{q_L} Y_{q_R}  \hspace{0.25mm} \alpha_1}{\pi } \, \ln \left( \frac{\mu_w^2}{\Lambda^2} \right) C_B (\Lambda) \,, \\[2mm]
C_\phi (\mu_w) & \simeq  - \frac{9 \hspace{0.25mm}  \alpha _2^2}{2} \, \ln \left(  \frac{\mu_w^2}{\Lambda^2} \right) C_W (\Lambda) \,,
\end{split}
\end{equation}
to the Wilson coefficients of the effective operators $O_y$ and $O_\phi$ at the scale $\mu_w$. Above $Y_{q_L}$~($Y_{q_R}$)  denote the hypercharges of the left-handed~(right-handed) quarks normalised such that $Y_{u_L} = Y_{d_L} = 1/6$, $Y_{u_R} = 2/3$ and $Y_{d_R} =-1/3$, while $\alpha_1$ and $\alpha_2$ are the coupling constants of $U(1)_Y$ and $SU(2)_L$, respectively. To the precision we are working at, the scale that enters the gauge couplings is undetermined. In~our numerical analysis we will employ $\alpha_1 \simeq 1/98$ and $\alpha_2 \simeq 1/29$ corresponding to weak scale values. Note that~(\ref{eq:7}) includes only contributions generated by the RG flow associated to $O_B$ and~$O_W$, while direct contributions from nonzero initial conditions~$C_k (\Lambda)$ of the remaining operators are neglected. Since in the second line of~(\ref{eq:7}) the  coupling constant $\alpha_2$ enters, one can expect a larger effect in direct detection from $O_W$ than from $O_B$.

In step $ii)$ the EW symmetry is spontaneously broken by the VEV of $\phi$ leading to mixing of $B_\mu$ and $W^3_\mu$ into the photon field~$A_\mu$ and massive quarks. For scales below~$\mu_w$ the operator basis thus contains
\begin{equation} \label{eq:8}
O_F = \bar \chi \chi \, F_{\mu \nu }  F^{\mu \nu} \,, \qquad 
O_q= m_q \, \bar \chi \chi \, \bar qq \,, 
\end{equation}
but no longer $O_B$, $O_W$ and $O_y$. In the normalisation~(\ref{eq:6}), the relevant tree-level matching conditions read 
\begin{equation} \label{eq:9}
\begin{split}
C_F (\mu_w) &= c_w^2  \hspace{0.25mm} C_B (\mu_w) + s_w^2  \hspace{0.25mm} C_W (\mu_w)  \,, \\[1mm]
C_q (\mu_w) &= C_y (\mu_w) - \frac{v^2}{m_h^2} \, C_\phi (\mu_w) \,,
\end{split}
\end{equation}
with $s_w$ and $c_w$ denoting the sine and cosine of the weak mixing angle. Our numerical results will utilise $s_w^2 \simeq 0.23$  and a Higgs boson mass of $m_h \simeq 125 \, {\rm GeV}$. 

Integrating out the top quark generates an effective interaction between DM and gluons of the form
\begin{equation} \label{eq:10}
O_G = \alpha_s \, \bar \chi \chi \, G_{\mu \nu}^a G^{a,\mu \nu} \,,
\end{equation}
with $\alpha_s$ given at the scale $\mu$ and $G_{\mu \nu}^a$ being the $SU(3)_c$ field strength tensor. At the one-loop level the corresponding Wilson coefficient is obtained via the Shifman-Vainshtein-Zakharov relation~\cite{Shifman:1978zn}. In terms of $C_t(\mu_w)$ as given in~(\ref{eq:9}), one finds 
\begin{equation} \label{eq:11}
C_G (\mu_w) = -\frac{1}{12 \pi} \, C_t (\mu_w)  \,,
\end{equation}
if the top quark is removed together with the Higgs and the EW gauge bosons as an active degree of freedom. 

The operator $O_q$ is defined to be invariant under QCD at the one-loop level. The RG evolution in step $iii)$ is, however, nontrivial since the operator $O_F$ mixes into~$O_q$ through the exchange of virtual photons. One obtains for scales $\mu$ with $\mu_w > \mu > m_q$ at LL order~\cite{Frandsen:2012db}
\begin{equation}  \label{eq:12}
C_q (\mu) \simeq C_q (\mu_w) +  \frac{3 \hspace{0.25mm} Q_q^2 \hspace{0.25mm} \alpha}{\pi } \,  \ln \left ( \frac{\mu^2}{\mu_w^2} \right ) C_F (\mu_w) \,,
\end{equation}
where  $Q_q$ is the electric charge of the corresponding quark. We will employ  the  value $\alpha \simeq 1/137$ for the electromagnetic coupling constant to obtain numerical predictions. 

At  the bottom and charm threshold one  has to integrate out the corresponding heavy quark by again applying~(\ref{eq:11}). Putting everything together and setting the scale $\mu_w$ equal to $m_W \simeq 80.4 \, {\rm GeV}$, we find for the relevant low-energy Wilson coefficients  $C_F (\mu_l)  \simeq C_F (\Lambda)$  and 
\begin{widetext}
\begin{equation} \label{eq:13}
\begin{split}
& \phantom{xxxxxxx} C_q (\mu_l)  \simeq \left( \frac{3 \hspace{0.25mm}  Y_{q_L} Y_{q_R} \hspace{0.25mm} \alpha_1}{\pi} \, C_B (\Lambda) + \frac{9 \hspace{0.25mm} \alpha _2^2}{2}  \frac{v^2}{m_h^2} \, C_W (\Lambda)   \right)\ln \left( \frac{m_W^2}{\Lambda^2} \right) + \frac{3 \hspace{0.25mm} Q_q^2 \hspace{0.25mm} \alpha}{\pi}  \hspace{0.25mm} C_F (\Lambda) \hspace{0.25mm} \ln \left( \frac{\mu_l^2}{m_W^2} \right) \,, \\[2mm]
& C_G (\mu_l) \simeq -\frac{1}{12 \pi} \,  \Bigg \{ \left( \frac{\alpha_1}{2 \hspace{0.25mm} \pi} \, C_B (\Lambda) + \frac{27 \hspace{0.25mm} \alpha _2^2}{2}  \frac{v^2}{m_h^2} \, C_W (\Lambda)   \right)\ln \left( \frac{m_W^2}{\Lambda^2} \right)  +\frac{\alpha}{3 \hspace{0.25mm} \pi} \, C_F (\Lambda) \left [ \ln \left ( \frac{m_b^2}{m_W^2} \right ) + 4  \ln \left ( \frac{m_c^2}{m_W^2} \right )  \right ] \Bigg \}  \,.
\end{split}
\end{equation}
\end{widetext}
Here $m_b \simeq 4.2 \, {\rm GeV}$, $m_c \simeq 1.3 \, {\rm GeV}$ and we have defined $C_F (\Lambda) = c_w^2  \hspace{0.25mm} C_B (\Lambda) + s_w^2  \hspace{0.25mm} C_W (\Lambda)$. Note that in the combination $C_F (\Lambda)$ the scale has been identified with $\Lambda$ although formally it should read $m_W$~$\big($see~(\ref{eq:9})$\big)$. This means that we ignore the self-mixing of $O_B$ and $O_W$ which is numerically negligible. Let us add that compared to~\cite{Frandsen:2012db},  the low-energy Wilson coefficients $C_q (\mu_l)$ and $C_G (\mu_l)$ contain additional contributions from~(\ref{eq:7}) and~(\ref{eq:9}), associated to EW symmetry breaking.

The expressions given in~(\ref{eq:13}) are LL accurate. Corrections beyond this order arise for instance from loop-level matching at the EW scale. A simple example of such an effect  is the one-loop matching correction to the Wilson coefficient of the operator $O_b$. This contribution is generated by a diagram similar to the one shown on the left-hand side in Fig.~\ref{fig:1} and involves  external bottom quarks as well as internal $W$ bosons and top quarks. Via~(\ref{eq:11}) the one-loop correction to the Wilson coefficient $C_b (\mu_w)$ then contributes to $C_G (\mu_l)$. Additional beyond-LL effects arise for example from the ${\cal O} (\alpha_s)$ corrections~\cite{Inami:1982xt} to the relation in~(\ref{eq:11}). We have explicitly verified that these higher-order contributions have only a minor effect~(at the level of 10\%) on the resulting DM-nucleon scattering cross section. In view of the hadronic uncertainties plaguing the calculation of  the direct detection rate, a computation of LL effects  thus seems sufficient. In our numerics we will hence employ the expressions for the low-energy Wilson coefficients as given in~(\ref{eq:13}). 

In step $iv)$ one has to evaluate the matrix elements of the effective operators $O_F$, $O_q$ and $O_G$ between nucleus states at the scale $\mu_l \simeq 1 \, {\rm GeV}$. Since in our  case next-to-leading order effects~\cite{Cirigliano:2012pq,Cirigliano:2013zta} play only a minor role, the spin-independent~(SI) cross section for elastic Dirac scattering on a nucleon simply reads~(cf.~\cite{Frandsen:2012db,Weiner:2012cb,Hill:2011be,Rajaraman:2011wf,Crivellin:2013ipa})
\begin{widetext}
\begin{equation} \label{eq:14}
\sigma _N^{\rm SI} \simeq \frac{m_{\rm red}^2 \hspace{0.25mm}  m_N^2}{\pi \hspace{0.25mm} \Lambda^6} \, \bigg | \hspace{0.25mm}  \frac{\alpha Z^2}{A} \hspace{0.25mm} f_F^N \hspace{0.25mm} C_F (\mu_l) + \sum_{q=u,d,s} f_q^N \hspace{0.25mm} C_q (\mu_l) - \frac{8 \hspace{0.25mm} \pi}{9} \hspace{0.25mm} f_G^N  \hspace{0.25mm} C_G (\mu_l) \hspace{0.25mm}  \bigg |^2 \,,
\end{equation}
\end{widetext}
where $m_{\rm red} = m_\chi \hspace{0.125mm} m_N/(m_\chi + m_N)$ denotes the reduced mass of the DM-nucleon system and $m_N \simeq 0.939 \, {\rm GeV}$ is the average nucleon mass. The term proportional to  $\alpha Z^2/A$ with $Z$ the total electric charge of the nucleus and $A$ its mass number stems from Rayleigh scattering of two photons on the entire nucleus. At zero-momentum transfer the corresponding  form factor reads $f_F^N \simeq 0.08, 0.10$ and $0.12$ for xenon, germanium and argon targets, respectively~\cite{Frandsen:2012db,Weiner:2012cb}.  For the remaining~$f_k^N$ describing the scalar couplings between the light and heavy quarks and the nucleon, we will employ the values $f_u^N \simeq 0.017$,  $f_d^N \simeq 0.036$ and $f_s^N \simeq 0.043$ from~\cite{Crivellin:2013ipa}. These form factors lead to $f_G^N = 1- \sum_{q=u,d,s} f_q^N \simeq 0.904$. To obtain the correct~$\sigma _N^{\rm SI}$ for Majorana DM, the right-hand side of~(\ref{eq:14})  has to be multiplied by a factor of 4.

\section{LHC bounds and relic density}
\label{sec:3}

As we have explained in the previous section, the effective interactions introduced in~(\ref{eq:4}) lead to a DM direct detection cross section via Higgs exchange diagrams involving gauge boson loops. In the case of Dirac fermions the induced partial Higgs decay width is given by 
\begin{equation} \label{eq:15}
\Gamma \left (h \to \bar \chi \chi \right ) = \frac{\big | C_\phi (m_h) \big |^2}{8 \pi}  \left ( \frac{v}{\Lambda} \right)^6 m_h \hspace{0.25mm} \beta_h^3  \,,
\end{equation}
where $C_\phi (m_h)$ is  the Wilson coefficient~(\ref{eq:7}) of  the operator~$O_\phi$ introduced in~(\ref{eq:5}) and $\beta_h = \left (1-4 \hspace{0.25mm} m_\chi^2/m_h^2 \right )^{1/2}$. For Majorana DM the rate~(\ref{eq:15}) is larger by a factor of~$2$. In terms of  $\Gamma \left (h \to \bar \chi \chi \right )$ and the total visible decay width $\Gamma ( h \to {\rm SM} )  \simeq 4.1 \cdot 10^{-3} \, {\rm GeV}$~\cite{CERNYellowReportPageBR3} of the SM Higgs with a mass of $125 \, {\rm GeV}$,  the invisible branching ratio  reads ${\rm Br} \left (h \to {\rm invisible} \right ) = \Gamma ( h \to \bar \chi \chi )/\sum_{f={\rm SM}, \bar \chi \chi} \Gamma ( h \to f)$. 

The latter quantity can be constrained by the  Higgs signal strength data obtained at the LHC and the Tevatron. Additional but weaker restrictions also follow from dedicated searches for $\slashed{E}_T$ signals arising from Higgs production \cite{Aad:2014iia,Chatrchyan:2014tja}. A very recent signal strength analysis~\cite{Cheung:2014noa} finds ${\rm Br} \left (h \to {\rm invisible} \right ) < 18.5 \%$ at 95\% confidence level~(CL) assuming that the Higgs has SM couplings but additional invisible decay modes.

Combining the above results the present bound on the invisible Higgs branching ratio provides only very loose constraints on the parameter space. Given the expected improvements with which the couplings of the Higgs can be extracted at future stages of the LHC and/or a possible ILC, relevant limits could however arise. In~\cite{ATL-PHYS-PUB-2013-014,CMS:2013xfa} the ATLAS and CMS experiments estimate  that at the $\sqrt{s} =14 \, {\rm TeV}$ LHC  with $3000 \, {\rm fb}^{-1}$  of integrated luminosity  a $95\%$~CL limit ${\rm Br} \left (h \to {\rm invisible} \right ) \lesssim 7\%$  may be obtainable. Assuming a NP scale of $\Lambda = 300 \, {\rm GeV}$, such a bound translates into the limit $C_W (\Lambda) \lesssim 1.1$ for Majorana DM with a mass of $10 \, {\rm GeV}$. For comparison, a $\sqrt{s} = 250 \, {\rm GeV}$ ILC with an integrated luminosity of~$250 \, {\rm fb}^{-1}$ can set a bound of ${\rm Br} \left (h \to {\rm invisible} \right ) \lesssim 0.9\%$ at $95\%$~CL~\cite{Asner:2013psa}. Such a precision would improve the above limit to $C_W (\Lambda) \lesssim 0.4$. 

Constraints on the operators~(\ref{eq:4}) also arise from collider data~\cite{Cotta:2012nj,Carpenter:2012rg,Nelson:2013pqa,Lopez:2014qja,Aad:2014vka}. We have updated these bounds by including the latest LHC searches for $h \to {\rm invisible}$ decays in vector boson fusion~\cite{Chatrchyan:2014tja}, a mono-photon~\cite{EXO-12-047-pas, ATLAS-CONF-2014-051}, a mono-$Z$~\cite{Aad:2014vka} and a~$\slashed{E}_T + W/Z \, (\to {\rm hadrons})$~\cite{Aad:2013oja} signal. In addition, we have considered the~$\slashed{E}_T + W \, (\to {\rm leptons})$ channel~\cite{ATLAS:2014wra,Khachatryan:2014tva} and  the newest mono-jet data~\cite{Khachatryan:2014rra}. While we reserve a detailed discussion of the constraints imposed by the individual search strategies for a future publication, it suffices to say, that depending on the choice of parameters, either the mono-photon or the mono-jet data give rise to the strongest restrictions at present. The 95\%~CL limits on the parameter space presented in the next section thus turn out to be more severe than the LHC limits found previously.

The effective operators $O_B$ and $O_W$ lead to velocity-suppressed DM annihilation rates. This means that writing the  non-relativistic annihilation cross sections as  $\sigma_k \hspace{0.25mm} v_\chi = a_k +b_k \hspace{0.25mm} v_\chi^2 + {\cal O} (v_\chi^4)$ with $v_\chi  \simeq 1.3 \cdot 10^{-3} \hspace{0.25mm} c$ the DM velocity, one has $a_B = a_W= 0$. The coefficients~$b_k$  on the other hand are nonzero. Employing the results given in~\cite{Chen:2013gya}, we find in the Dirac case 
\begin{widetext}
\begin{equation} \label{eq:16}
\begin{split}
& \phantom{xxxxxxxxxxxx} b_B = \frac{\big | C_B(\Lambda) \big |^2}{ \pi} \,  \frac{m_\chi^4}{\Lambda^6} \left [ c_w^4 + \frac{c_w^2 s_w^2}{8} \hspace{0.25mm} \beta_{\gamma Z}^2 \left ( x_Z - 4 \right)^2  + \frac{s_w^4}{8} \hspace{0.25mm} \beta_{Z Z} \left ( 3 x_Z^2 - 8 x_Z + 8  \right)  \right ]  \,, \\[2mm]
& b_W  = \frac{\big | C_W(\Lambda) \big |^2}{ \pi} \,  \frac{m_\chi^4}{\Lambda^6} \left [ s_w^4 + \frac{c_w^2 s_w^2}{8} \hspace{0.25mm} \beta_{\gamma Z}^2 \left ( x_Z - 4 \right)^2  + \frac{c_w^4}{8} \hspace{0.25mm} \beta_{Z Z} \left ( 3 x_Z^2 - 8 x_Z + 8  \right)+ \frac{1}{4} \hspace{0.25mm} \beta_{W W} \left ( 3 x_W^2 - 8 x_W + 8  \right) \right ]\,, 
\end{split}
\end{equation}
\end{widetext}
where we have defined $\beta_{kl} =  \left (1- (m_k + m_l)^2/(4 \hspace{0.25mm} m_\chi^2) \right)^{1/2}$ and $x_k = m_k^2/m_\chi^2$. Hereafter we will use $m_Z \simeq 91.2 \, {\rm GeV}$. Note that the coefficients in (\ref{eq:16}) are larger by a factor of 4 if DM is Majorana.  Using the results for $b_B$ and $b_W$ the relic density can then be calculated using~\cite{Kolb:1990vq}
\begin{equation} \label{eq:17}
\Omega_\chi h^2 \simeq  0.11 \; \frac{1.58 \cdot 10^{-8} \, {\rm GeV}^{-2} \;  s_\chi}{b_B+b_W} \,, 
\end{equation}
where $s_\chi = 2$ or $1$ in the case of Dirac or Majorana fermions. The cross section for DM annihilation into~$\gamma \gamma$ can be obtained from the above results by employing $( \sigma \hspace{0.25mm} v_\chi )_{\gamma \gamma} \simeq \left (b_B + b_W \right )/s_\chi \, v_\chi^2  \big |_{\beta_{kl}=0}$.

\section{Numerical analysis}
\label{sec:4}

\begin{figure}
\begin{center}
\includegraphics[width=55.5ex]{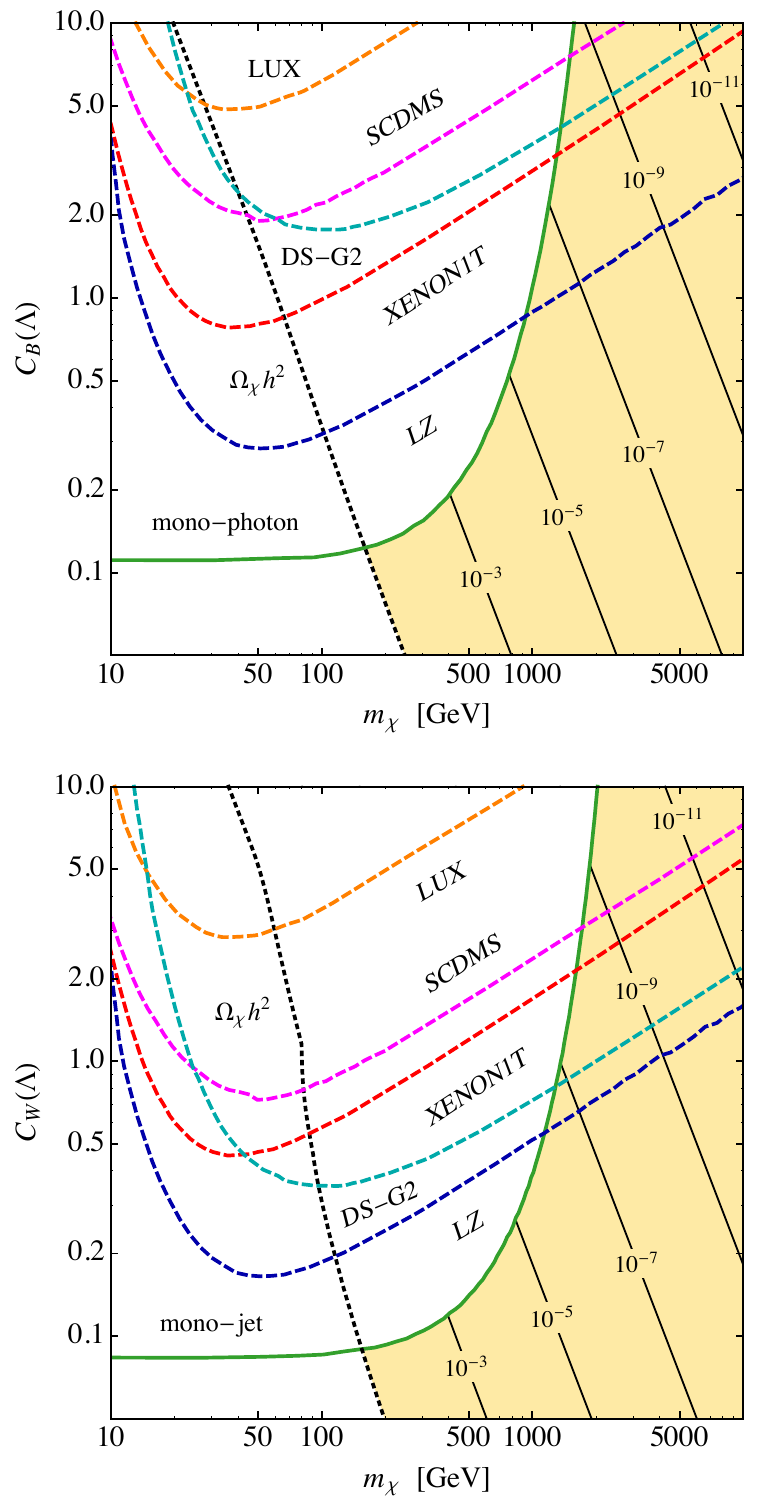}
\end{center}
\vspace{-2mm}
\caption{\label{fig:2} Restrictions in the $m_\chi$--$\hspace{0.25mm} C_B (\Lambda)$~(upper panel) and $m_\chi$--$\hspace{0.25mm} C_W (\Lambda)$~(lower panel) plane, assuming DM to be Majorana and setting $\Lambda = 300 \, {\rm GeV}$. The green curves illustrate the best limits from $\slashed{E}_T$ searches at the LHC, while the black dotted  lines correspond to the observed value $\Omega_\chi h^2 = 0.11$ of the relic density. The coloured dashed  curves mark the bounds from existing and future direct detections experiments. The currently allowed parameter regions are indicated by yellow shading. The contour lines denote the fraction of the observed relic density obtained from the operator under consideration. For Dirac DM the bounds on the Wilson coefficients~$C_B (\Lambda)$ and  $C_W (\Lambda)$ from the LHC, the relic density and direct detection are weaker by a factor of $\sqrt{2}$,  $2 \sqrt{2}$ and  $2$.}
\end{figure}

In the following we will compare the bounds from direct detection originating from the loop effects discussed in detail in Sec.~\ref{sec:2} to the limits from $\slashed{E}_T$ searches at colliders  and the requirement that the correct relic density is reproduced. We will present results for real Wilson coefficients, fixing the NP scale $\Lambda$ to  $300 \, {\rm GeV}$ and allowing either $C_B (\Lambda)$ or $C_W (\Lambda)$ to be nonzero. These two scenarios will be called I and II below. In principle, it is also possible that additional operators affect direct detection both constructively as well as destructively. Therefore, we assume the absence of tuning and let the operators~(\ref{eq:4}) to saturate the experimentally allowed cross section, considering a single interaction at a time. Furthermore, one must keep in mind that our effective field theory (EFT) approach does not necessarily hold when applied to LHC searches. However, the applicability of the EFT depends on the way the effective operators are generated, i.e.~on the details of the ultraviolet (UV) complete theory. See~\cite{Liu:2013gba} for a study of LHC signatures that does not rely on the EFT description. Making the above statements precise would require to study a concrete UV completion, which is beyond the scope of the present work. 

For  Majorana fermions the various constraints in the $m_\chi$--$\hspace{0.25mm} C_B (\Lambda)$~(upper panel) and $m_\chi$--$\hspace{0.25mm} C_W (\Lambda)$~(lower panel) plane are shown in Fig.~\ref{fig:2}. In scenario~I, we find that the recent mono-photon search by CMS~\cite{EXO-12-047-pas} represents the most stringent LHC bound. As illustrated by the green curve, this constraint requires $C_B (\Lambda) \lesssim 0.12$ for $m_\chi \lesssim 100 \, {\rm GeV}$.  The parameter space to the left (right) of the dashed back line corresponds to $\Omega_\chi \hspace{0.25mm} h^2 > 0.11$ ($\Omega_\chi \hspace{0.25mm} h^2 < 0.11$), assuming standard freeze-out of symmetric DM. In regions where DM underproduction is predicted, the observed relic abundance $\Omega_\chi \hspace{0.25mm} h^2 \simeq 0.11$ \cite{Hinshaw:2012aka} can still be obtained if an initial asymmetry is present in the dark sector in analogy to the baryon asymmetry. To calculate bounds from direct detection in these regions, we assume that such a mechanism does indeed increase the abundance to account for all of the observed DM. From the intersection of the mono-photon bound and the relic constraint, we obtain a lower bound on the DM mass of~$m_\chi \gtrsim 160 \,{\rm GeV}$.  The restrictions  imposed by direct detection are indicated by the dashed coloured curves. One observes that already the first  LUX results~\cite{Akerib:2013tjd} (orange) lead to a, though admittedly weak constraint in the  $m_\chi$--$\hspace{0.25mm} C_B (\Lambda)$ plane. Next-generation detectors such as SuperCDMS~(SCDMS)~\cite{Brink:2012zza}~(magenta),  XENON1T~\cite{Aprile:2012zx}~(red) and  DarkSide-G2 (DS-G2)~\cite{Wright:2011pa}~(cyan)  will however be able  to improve on the existing bounds, and the sensitivity of the proposed LUX-ZEPLIN~(LZ) DM search experiment~\cite{Malling:2011va}~(blue) will allow to probe unconstrained values of~$C_B (\Lambda)$ for~$m_\chi \gtrsim 930 \, {\rm GeV}$. 

In scenario~II, we find that the latest mono-jet data from CMS~\cite{Khachatryan:2014rra}~(green curve) provide the strongest collider constraint at the moment. In fact, we obtain~$C_W (\Lambda) \lesssim 0.09$ for DM Majorana masses below about $100 \, {\rm GeV}$. Note that the latter bound significantly exceeds the limit that one might be able to extract from a determination of ${\rm Br} \left (h \to {\rm invisible}\right)$ at an ILC. Combining the LHC and the DM abundance (dotted black line) constraints, we can again derive a lower limit on the DM mass and obtain $m_\chi \gtrsim 150 \,{\rm GeV}$. Turning to the direct detection limits (dashed coloured curves), we observe that in scenario~II the existing LUX data~(orange) are also not sufficient to test unexplored parts of the parameter space. However, the anticipated sensitivity of SCDMS~(magenta), XENON1T~(red), DS-G2~(cyan) and~LZ~(blue) will again allow to set  novel upper bounds on $C_W (\Lambda)$, if DM  is sufficiently heavy. 

Note that for fixed $\Lambda$ and $m_\chi$ the direct detection constraints on $C_W (\Lambda)$ are always stronger than those on~$C_B (\Lambda)$.  This feature can be traced back to the destructive interference between the Rayleigh contribution associated to $O_F$ and  the corrections from $O_q$ and $O_G$ arising due to operator mixing and threshold corrections (this interference effect has already been noticed in~\cite{Frandsen:2012db}). In scenario~I, we find that the term proportional to $C_F (\mu_l)$ in~(\ref{eq:14}) always dominates over the remaining corrections. The dominance of the Rayleigh contribution has two origins. First, in~(\ref{eq:9}) the Wilson coefficient~$C_B (\mu_w)$ enters~$C_F (\mu_w)$ with a factor of $c_w^2$ and second, the numerical coefficients in~(\ref{eq:13}) multiplying~$C_B(\Lambda)$ are much smaller than those in front of the Wilson coefficients  $C_W(\Lambda)$.  In scenario~II, on the other hand, the logarithmically-enhanced effects entering via~(\ref{eq:7}) and (\ref{eq:9}) dominate in (\ref{eq:14}), while Rayleigh scattering  provides only a minor contribution to $\sigma_N^{\rm SI}$. Note that the aforementioned destructive interference is weaker for heavier nuclei since the Rayleigh contribution to the SI DM-nucleon cross section scales with $\alpha Z^2/A$. This property explains why the DS-G2 constraint  is, compared to XENON1T, less powerful in scenario I, while in scenario~II the DS-G2 experiment will be able to probe a larger part of the parameter space than XENON1T. 

We finally add that the  annihilation cross sections into  photon pairs corresponding to the allowed parameter space shown in Fig.~\ref{fig:2} (yellow shaded regions) lie in the range of around $[10^{-42}, 10^{-30} ] \, {\rm cm}^{3} \, {\rm s}^{-1}$ $\big($$[10^{-43}, 10^{-31}] \, {\rm cm}^{3} \, {\rm s}^{-1}$$\big)$ in scenario I~(II). Even order-of-magnitude improvements of the present-day indirect detection limits by $\gamma$-ray detector specifically designed for DM detection (such as the Dark Matter Array described in~\cite{Bergstrom:2010gh}) will thus not provide any relevant constraint on the interactions (\ref{eq:4}). 

\section{Conclusions}
\label{sec:5}

In this work we have highlighted the importance of a systematic analysis of loop effects induced by SM fields to connect effective operators generated at the NP scale~$\Lambda$ with direct detection rates that probe the DM-SM interactions at low energies. We considered the case of a fermionic DM particle which is a singlet under the SM gauge group. Focusing on the only two dimension-7 operators that involve regular EW field strength tensors and lead to velocity-suppressed DM annihilation rates into photon pairs, we identified new contributions to DM-nucleon scattering that arise from gauge boson loops in the unbroken $SU(2)_L \times U(1)_Y$ theory. The Wilson coefficients of the considered  effective DM-gauge boson interactions have previously been bounded by LHC data, but the limits that originate from loop-induced direct detection have not been studied. Our calculations enable us to perform such an analysis incorporating all relevant LL-enhanced effects. While it turns out that the current sensitivity of direct detection experiments is insufficient to compete with the existing LHC constraints, the next generation of shielded underground detectors will be able to probe unexplored parts of the parameter space. Unlike the LHC, direct detection experiments are not kinematically limited, so that for DM masses of a few ${\rm TeV}$ the latter search strategy may, for some time, provide  the only window on the  interactions considered here. 

\begin{acknowledgments}
We are grateful to M.~Gorbahn and  F.~Kahlhoefer for helpful discussions and thank the latter as well as L.~Tunstall for useful comments on the manuscript. AC is supported by a Marie~Curie Intra-European Fellowship of the European Community's $7^{th}$ Framework Programme under contract number PIEF-GA-2012-326948. He is also grateful to the Mainz Institute for Theoretical Physics~(MITP) of the DFG cluster of excellence ``Precision Physics, Fundamental Interactions and Structure of Matter'' for hospitality and partial support during the completion of this work. UH  acknowledges  the hospitality and support of the CERN theory division and the Munich Institute for Astro- and Particle Physics~(MIAPP) of the DFG cluster of excellence ``Origin and Structure of the Universe''.
\end{acknowledgments}

\end{document}